\documentclass{mc2015}	

\usepackage[T1]{fontenc}         % Use T1 encoding instead of OT1
\usepackage[utf8]{inputenc}      % Use UTF8 input encoding
\usepackage{microtype}           % Improve typography
\usepackage{booktabs}            % Publication quality tables
\usepackage{amsmath}
\usepackage{graphicx}
\usepackage{float}
\usepackage[exponent-product=\cdot]{siunitx}
\usepackage[colorlinks,breaklinks]{hyperref}
\hypersetup{linkcolor=black, citecolor=black, urlcolor=black}

\usepackage{lipsum}
\usepackage{amssymb}

\newcommand{\Sn}{\ensuremath{S_N}}
\newcommand{\Macro}{\ensuremath{\Sigma}}

\newcommand{\ve}[1]{\ensuremath{\mathbf{#1}}}

\authorHead{R.N.\ Slaybaugh, T.M.\ Evans, G.G.\ Davidson, et al.}
\shortTitle{RQI and MGE for Neutron Transport}

\begin{document}

\title{Rayleigh Quotient Iteration with a Multigrid in Energy Preconditioner for Massively Parallel Neutron Transport}

\author{R.N.\ Slaybaugh}
\affil{
  Department of Nuclear Engineering\\
  University of California, Berkeley\\
  4173 Etcheverry Hall, Berkeley, CA 94720\\
  slaybaugh@berkeley.edu
}
\author{T.M.\ Evans}
\author{G.G.\ Davidson}%\footnote{Footnote, if necessary}}
\affil{Radiation Transport Group\\
  Oak Ridge National Laboratory, P.O. Box 2008, Oak Ridge, TN 37831\\
  evanstm@ornl.gov; davidsongg@ornl.gov}

\author{P.P.H.\ Wilson}
\affil{
  Department of Nuclear Engineering and Engineering Physics\\
  University of Wisconsin -- Madison\\
  419 ERB, 1500 Engineering Drive, Madison, WI 52706\\
  wilsonp@cae.wisc.edu
}

\maketitle

\begin{abstract}
Three complementary methods have been implemented in the code Denovo  that accelerate neutral particle transport calculations with methods that use leadership-class computers fully and effectively: a multigroup block (MG) Krylov solver, a Rayleigh quotient iteration (RQI) eigenvalue solver, and a multigrid in energy preconditioner. The multigroup Krylov solver converges more quickly than Gauss Seidel and enables energy decomposition such that Denovo can scale to hundreds of thousands of cores. The new multigrid in energy preconditioner reduces iteration count for many problem types and takes advantage of the new energy decomposition such that it can scale efficiently. These two tools are useful on their own, but together they enable the RQI eigenvalue solver to work. Each individual method has been described before, but this is the first time they have been demonstrated to work together effectively.  

RQI should converge in fewer iterations than power iteration (PI) for large and challenging problems. RQI creates shifted systems that would not be tractable without the MG Krylov solver. It also creates ill-conditioned matrices that cannot converge without the multigrid in energy preconditioner. Using these methods together, RQI converged in fewer iterations and in less time than all PI calculations for a full pressurized water reactor core. It also scaled reasonably well out to 275,968 cores.

\end{abstract}

%%%%%%%%%%%%%%%%%%%%%%%%%%%%%%%%%%%%%%%%%%%%%%%%%%%%%%%%%%%%%%%%%%%%%
\section{Introduction}\label{sec:intro}

The goal of this research is to accelerate neutral particle transport calculations with methods that use large computers fully and effectively, facilitating the design of better nuclear systems. Typical steady-state deterministic transport problems today are three-dimensional, have up to thousands $\times$ thousands $\times$ thousands of mesh points, use up to $\sim$150 energy groups, include accurate expansions of scattering terms, and are solved over many angular directions. The next generation of challenging problems are even more highly refined and the flux must be calculated quickly and accurately. Very large computers, like Titan~\cite{Titan2013}, are now available to perform such high-fidelity calculations. New solution methods must outperform existing ones that are not able to take full advantage of new computer architectures or have convergence properties that limit their usefulness for difficult problems. 

Three complementary methods have been implemented in the code Denovo \cite{Evans2009d} that accomplish this goal: a multigroup block (MG) Krylov solver, a Rayleigh quotient iteration (RQI) eigenvalue solver, and a multigrid in energy (MGE) preconditioner. Each individual method has been described before (see \cite{Davidson2013}, \cite{Slaybaugh2012}, \cite{Slaybaugh2013}), but this is the first time they have been demonstrated to work together effectively.

The MG Krylov solver was designed to improve performance when compared to Gauss Seidel iteration (GS) and to dramatically increase the number cores Denovo can use. Instead of sequentially solving each group with some inner iteration method and then using GS for outer iterations to converge the upscattering (which can be time consuming when many groups contain upscattering), the MG Krylov solver treats a block of groups at once such that the inner-outer iteration structure is removed. This results in faster solutions for most problem types. In addition, the block Krylov solver allows energy groups to be solved simultaneously because the multigroup-sized matrix-vector multiply can be divided up in energy and parallelized--extending the number of cores that can be used efficiently by Denovo from tens of thousands to hundreds of thousands. 

An MGE preconditioner was added to Denovo to reduce iteration count for all problem types and to address convergence issues associated with RQI. The preconditioner uses a multigrid method in the energy dimension. A set of energy grids with increasingly coarse energy group structures are created. This is implemented in a way that easily and efficiently takes advantage of the new energy decomposition. The multigrid algorithm is applied within each energy set such that the energy groups are only restricted and prolonged between groups on that set. Inter-set communication in the preconditioner is kept low, so the scaling in energy is very good. 

Theory indicates that RQI should converge in fewer iterations than traditional eigenvalue solvers like power iteration (PI), particularly for problems that are challenging for those solvers. However, RQI causes the systems it operates on to become nearly singular, and thus, it requires a preconditioner. Further, the implementation of RQI results in a set of equations that is mathematically equivalent to having upscattering in every group, so the scattering matrix becomes energy-block dense. Handling energy-block dense systems when there are more than a few energy groups is not tractable with GS as the multigroup solver. It is only the MG Krylov solver that makes RQI reasonable to use when there are many energy groups.

The remainder of this paper details why these methods are so complementary and presents results demonstrating this behavior. Section \ref{sec:background} discusses each of the new methods in the context of commonly-used methods. Section \ref{sec:pastwork} gives an overview of relevant past work. New results from using the three new methods together are shown in Section \ref{sec:results}, and concluding remarks are made in Section \ref{sec:conclusions}.

%%---------------------------------------------------------------------------%%
%%---------------------------------------------------------------------------%%
%%---------------------------------------------------------------------------%%
\section{Background}\label{sec:background}

The eigenvalue form of the multigroup \Sn\ equations can be written in operator form as
\begin{equation}
  \ve{L}\psi = \ve{MS}\phi + \frac{1}{k}\ve{M}\chi \ve{f}^{T}\phi \:. \label{eq:eigenvalue}
\end{equation}
Here $\ve{L}$ is the first-order linear differential transport operator; $\ve{M}$ is the moment-to-discrete operator that projects the angular flux moments, $\phi$, onto discrete angles; $\ve{S}$ is the scattering matrix; $\ve{f}$ contains the fission source, $\nu \Macro_{f}$; and $k$ is the asymptotic ratio of the number of neutrons in one generation to the number in the next. This can be converted to a fixed-source equation by replacing the fission term with an external source, $q$. The angular flux moments are related to the angular flux through the discrete-to-moment operator: $\phi = \mathbf{D} \psi \:$. Using this relationship, Eq.\ \eqref{eq:eigenvalue} can be rearranged such that it is a function of only $\phi$. The formulation is aided by defining $\ve{T} = \ve{DL}^{-1}$ and $\ve{F} = \chi \ve{f}^{T}$ \cite{Evans2010}:
\begin{equation}
  (\ve{I} - \ve{TMS})\phi = \frac{1}{k} \ve{TMF} \phi \:. \label{eq:OperatorEvalForm}
\end{equation}

Once the matrices are multiplied together, a series of single ``within-group'' equations that are each only a function of space and angle result. If the groups are coupled together by neutrons scattering from a low energy group to a higher energy group, then iterative ``multigroup'' solves over the coupled portion of the energy range may be required. If the eigenvalue is desired, an additional ``eigenvalue'' solve is needed. 

%%---------------------------------------------------------------------------%%
%%---------------------------------------------------------------------------%%
\subsection{Block Krylov Solver}\label{sec:blockkrylov}

Traditionally, the multigroup solve has been done with Gauss Seidel iteration. A space-angle solve using a within-group solver, such as source iteration or a Krylov method, is performed for each energy group in series. The groups are solved from $g=0$, the highest energy, to $g=G$, the lowest. For a group $g$ and an energy iteration index $j$ this is \cite{Evans2010}
\begin{equation}
  \bigl( \ve{I} - \ve{TMS}_{gg} \bigr) \phi^{j+1}_{g} = \ve{TM} \bigl( \sum_{g'=0}^{g-1}\ve{S}_{gg'}\phi^{j+1}_{g'} + \sum_{g'=g+1}^{G} \ve{S}_{gg'}\phi^{j}_{g'}  + q_{g} \bigr)  \:.
 \label{eq:up-GS}
\end{equation}
The first term on the right includes downscattering contributions from higher energies, and the second term represents upscattering contributions from lower energy groups that have not yet been converged for this energy iteration. 
%Groups that only contain downscattering are simply solved once since the second term on the right is zero. Groups with upscattering, however, must be iterated until they converge. 
Convergence of GS is governed by the spectral radius of the system, so the method can be very slow when upscattering has a large influence on the solution \cite{Adams2002}. GS is fundamentally serial in energy because of how the group-to-group coupling is treated. 

The MG Krylov solver removes the traditional within-group, multigroup iteration structure to make one space-angle-energy iteration level. This allows the solver to handle upscattering efficiently since Krylov methods generally converge more quickly than GS for problems with upscattering, and enables parallelization in the energy dimension. The solver has been shown to successfully scale to hundreds of thousands of cores because of the addition of energy parallelization \cite{Davidson2013,Slaybaugh2011}. 

The multigroup Krylov method applied to a block of groups is shown here, where $\ve{S}_{\text{block}}$ contains the block of upscattering groups and $\ve{S}_{\text{block\_source}}$ has the downscattering-only groups:
\begin{equation}
  \underbrace{(\ve{I} - \ve{TMS}_{\text{block}})}_{\tilde{\ve{A}}}\phi_{\text{block}}^{n+1} = \ve{TM}(\ve{S}_{\text{block\_source}}\phi_{\text{block\_source}}^{n+1} + q) \:.
  \label{eq:MGkrylov}
\end{equation}
The AztecOO package of Trilinos \cite{1089021} provides Denovo's Krylov solver, with a choice of either GMRES($m$) or BiCGSTAB. The Krylov solver is given an operator that implements the action of $\ve{\tilde{A}}$, or the matrix-vector multiply and sweep. In the MG Krylov solver, $\ve{\tilde{A}}$ is applied to an iteration vector, $v$, containing the entire block of groups instead of just one group.

To implement the energy parallelization, the problem is divided into energy sets, with groups distributed evenly among sets. After each set performs its part of the matrix-vector multiply, a global reduce-plus-scatter is the only required inter-set communication. The added energy decomposition offers the ability to further decompose a problem, even if the performance limit of spatial decomposition has been reached. The total number of cores is equal to the number of computational domains, that is, the product of the number of energy sets and the number of spatial blocks. For 20,000 spatial blocks and 10 energy sets, which is a reasonable decomposition, 200,000 cores will be used. See Ref.\ \cite{Davidson2013}, \cite{Slaybaugh2011}, or \cite{Evans2010}  for more details. It is worth noting that the scaling limits seen in Denovo are not fundamental; good scaling beyond tens of thousands of cores is possible without energy decomposition (see, e.g., \cite{Bailey2014}). 

%%---------------------------------------------------------------------------%%
%%---------------------------------------------------------------------------%%
\subsection{Eigenvalue Solvers}
\label{sec:eigenvalue}
A common way to solve $k$-eigenvalue problems is with power iteration. This method is attractive because it only requires matrix-vector products and two vectors of storage space. 
\begin{align}
  \ve{A}\phi &= k\phi \:; \qquad \ve{A} = (\ve{I} - \ve{TMS})^{-1} \ve{TMF} \:,\label{eq:EnergyDepEval} \\
  \phi^{i+1} &= \frac{1}{k^i}\ve{A}\phi^{i} \:; \qquad 
  k^{i+1} = k^i \frac{\ve{f}^T \phi^{i+1}}{\ve{f}^T \phi^i} \:.
  \label{eq:PowerIteration}
\end{align} 
PI uses the form of the problem seen in Eq.\ \eqref{eq:EnergyDepEval} and then iterates as shown in Eq.\ \eqref{eq:PowerIteration}, where $i$ is the iteration index. 
%This converts the generalized form of the eigenvalue problem seen in Eq.\ \eqref{eq:OperatorEvalForm} to the ordinary form. In the generalized form, the eigenvector-value pair is $(\phi, \frac{1}{k})$, and in the ordinary form it is $(\phi, k)$. 
%In legacy applications, the eigenvector is often the fission source rather than the flux moments \cite{Evans2011}, \cite{Lewis1993}.
Inside of PI, the application of $\ve{A}$ to $\phi$ requires the solution of a multigroup problem that looks like a fixed source problem: $(\ve{I} - \ve{TMS})y^{i} = \ve{TMF}\phi^{i}$.
PI's convergence can be very slow for problems of interest. 
For an $n \times n$ matrix $\ve{A}$, an eigenvalue-vector pair satisfies $\ve{A}x_{i} = \lambda_{i}x_{i}$ for $i = 1,...,n$. 
Let $\sigma(\ve{A}) \equiv \{\lambda \in \mathbb{C} : rank(\ve{A} - \lambda \ve{I}) < n\}$ 
be the spectrum of $\ve{A}$ and the eigenvalues be ordered as $|\lambda_{1}| > |\lambda_{2}| \ge \dots \ge |\lambda_{n}| \ge 0$. 
The error from PI is reduced in each iteration by a factor of $\ve{A}$'s dominance ratio, $\lambda_{2}/\lambda_{1}$. 
PI will converge slowly for loosely coupled systems because $\lambda_{2}$ is close to $\lambda_{1}$ \cite{Trefethen1997}. 

Shifted inverse iteration (SII) typically converges more quickly than PI. SII capitalizes on the fact that for some shift $\mu$ and some matrix $\ve{P}$, $(\ve{P} -\mu \ve{I})$ will have the same eigenvectors as $\ve{P}$. If $\mu\notin \sigma(\ve{P})$, %where $\sigma(\ve{A})$ is the spectrum of of $\ve{A}$, 
then $(\ve{P} - \mu \ve{I})$ is invertible and $\sigma([\ve{P} - \mu \ve{I}]^{-1}) = \{1/(\lambda - \mu):\lambda \in \sigma(\ve{P})\}$. Eigenvalues of $\ve{P}$ that are near the shift will be transformed to extremal eigenvalues that are well separated from the others. The shifted and inverted matrix is used in a power iteration-type scheme. Given a good shift, $\mu \approx \lambda_1$, SII usually converges more quickly than PI, especially for loosely coupled systems \cite{Trefethen1997}.%, \cite{Allen2002}.

Rayleigh Quotient Iteration is an SII method that uses an optimal shift: the Rayleigh quotient (RQ). For a generalized eigenvalue problem $\beta \ve{A}x = \alpha \ve{B}x$, the RQ is
\begin{equation}
  \rho = \frac{y^{T} \ve{A} x}{y^{T} \ve{B} x} \:.
  \label{eq:RQ}
\end{equation}
If $x$ and $y$ are right and left eigenvectors corresponding to the scalars $\alpha$ and $\beta$, respectively, then $\alpha = y^{T} \ve{A} x$ and $\beta =  y^{T} \ve{B} x$. In this case the system's eigenvalue is $\gamma = \frac{\alpha}{\beta}$ and $\rho = \gamma$ \cite{Stewart2001}. The RQ provides the minimum residual for the eigenvalue problem in the least squares sense. %, where equivalence holds only when $\tau = \rho$ and $x \ne 0$:
%
%\begin{equation}
%  ||(\ve{A} - \tau\ve{B})x||^{2} \ge ||\ve{A}x||^{2} - ||\rho\ve{B}x||^{2} \:.
%\end{equation}
%
The Rayleigh quotient is thus an optimal guess for an eigenvalue given a vector close to the corresponding eigenvector. 
The idea of RQI is to use the RQ as the shift in SII, but the shift is updated with the new eigenvector estimate at every iteration $i$: $\mu_i = \gamma_{i-1} - \rho_i$.
RQI has better convergence properties than SII since the RQ is an optimal guess for the eigenvalue of interest. For more details on the Rayleigh quotient and RQI in general, refer to Ref.\ \cite{Parlett1974}.

RQI has been implemented in Denovo, as detailed in Ref.\ \cite{Slaybaugh2012}, by subtracting $\rho \ve{TMF}$ from both sides of Eq.\ \eqref{eq:OperatorEvalForm}. This gives the following shifted system, where $\gamma \equiv \frac{1}{k}$:
\begin{equation}
  (\ve{I} - \ve{TM}\ve{\tilde{S}})\phi =( \gamma - \rho) \ve{TMF} \phi  \:, 
  \label{eq:OperatorShiftedEval} 
\end{equation}
and $\ve{\tilde{S}} \equiv \ve{S} + \rho\ve{F}$. This new matrix, $\ve{\tilde{S}}$, is energy-block dense since the fission matrix is dense and looks like one big upscattering block. Traditional solution methods for the fixed source part of the equation do not handle dense scattering matrices well. % This has hampered the implementation of SII in multigroup, 3-D codes because solving many groups that have upper-triangular scattering entries can be time-prohibitive. 
The RQI solver added to Denovo uses the MG Krylov solver, designed to handle dense scattering matrices effectively, to overcome the use of a dense $\ve{\tilde{S}}$. In addition, RQI can be decomposed in energy and takes advantage of the scaling properties of the multigroup solver. 

Using RQI in combination with a Krylov method, however, raises some concerns about whether or not it can converge because the matrix becomes so ill-conditioned. Peters and Wilkinson \cite{Peters1979}, however, proved that ill-conditioning is not a fundamental problem for inverse iteration methods. Trefethen and Bau \cite{Trefethen1997} assert that this is the case as long as the fixed source portion is solved with a backwards stable algorithm. Paige et al.\ \cite{Paige2006} demonstrated that GMRES is backwards stable when finding $x$ in $\ve{A}x = b$ for a ``sufficiently nonsingular $\ve{A}$'' and define associated criteria. For ill-conditioned systems, then, Krylov methods may not be backwards stable and will tend to converge very slowly. Many researchers have found that Krylov methods must be preconditioned to achieve good convergence in practice, e.g.\ \cite{Trefethen1997}, \cite{Paige2006}. The past work section below demonstrates that RQI \textit{does} need preconditioning for convergence in problems of interest.

\subsection{Multigrid in Energy Preconditioner}
\label{sec:precond}
Preconditioning is needed to make the RQI algorithm tractable and is important for increasing the robustness of Krylov methods and decreasing Krylov iteration count. This is particularly true for the multigroup Krylov solver; it can create large subspaces because it forms the subspaces with multiple-group-sized vectors. As a result, any reduction in iteration count will be of significant benefit in terms of memory and cost per iteration. A right preconditioner that does multigrid in the energy dimension, meaning the grids are in energy such that the energy group structure is coarsened and each lower grid has fewer groups, and is designed to work with the MG Krylov solver was implemented in Denovo \cite{Slaybaugh2013}. To understand why multigrid in energy makes sense for neutron transport, some highlights about these methods are discussed here (see Ref.\ \cite{Briggs2000} for details). 

The error in $x_i$, the $i$th guess for $\ve{A}x_i=b_i$, can be written as a combination of Fourier modes. Each Fourier mode has a frequency; low-frequency modes are smooth and high-frequency modes are oscillatory. Stationary iterative methods remove high-frequency error components quickly but can take many iterations to remove the low-frequency ones. Multigrid methods use the smoothing effects of iterative methods by making smooth errors look oscillatory and thus easier to remove. These methods remove the low-frequency error modes by restricting error to coarser grids, removing now higher-frequency error, and prolonging the smoothed error back to the fine grid. 

%The multigrid in energy method is implemented as a right preconditioner in Denovo, the problem is broken into two steps:
%%
%\begin{enumerate}
%  \item With a Krylov method solve \hspace*{1 em}$\ve{AG}^{-1}y = b$, 
%  \item after finding $y$, calculate \hspace*{1 em}$\phi = \ve{G}^{-1}y$.
%\end{enumerate}
%%
%Here $\ve{G}^{-1}$ represents the application of the preconditioner, and $y$ is defined as $\ve{G}\phi$; recall that $\ve{A} = \ve{I} - \ve{TMS}$.  
An important principle is that the preconditioner is only attempting to approximately invert $\ve{A}$. It is therefore reasonable to use a less accurate angular discretization in the preconditioner than the rest of the code. For example, the whole problem may be solved at $S_{10}$, but the preconditioner could use $S_{2}$. %The user can specify an angular quadrature set to use in the preconditioner that is different from the angular quadrature set used in the rest of the problem; the default is to use the same set in both. At this time, this option has only been implemented for vacuum boundary conditions. 
% The finest grid is the input energy structure, and the coarsest grid has one or a few groups. Each level has half as many groups as the previous level, rounded up if applicable. If there are $G+1$ groups on the fine grid there will be either $\frac{G+1}{2}$ or $\frac{G+2}{2}$ groups on the coarse grid. This is conceptually straightforward because the energy groups can be combined (restricted) and separated (prolonged) linearly. For details on the implementation of the preconditioner consult Ref. \cite{Slaybaugh2013}
%
%The implemented restriction operator is a simple averaging scheme. Neighboring fine data are averaged together to make coarse data. To prolong from a coarse to a fine grid, the points that line up between the grids are mapped directly. To fill in the intermediate points on the fine grid, the adjacent coarse values are averaged. There are other restriction and prolongation operators that are more rigorous and would preserve more accuracy when transferring between grids than those implemented. However, the implemented methods were found to be sufficient in practice.

The user chooses the number of V-cycles done for each preconditioner application. One V-cycle proceeds from the finest grid to the coarsest grid and back to the finest. Each additional V-cycle should remove more error, but has a computational cost. The depth of the V-cycle can also be specified by the user. The default behavior is determined by the number of groups, such that the grids will be coarsened until there is only one energy group. The number of grids needed in that case is $\text{floor}\bigl( \log_{2}(G-1) \bigr) + 2$ \cite{Slaybaugh2013}.

Some number of relaxations are performed on each level while traversing down and up the grids in a V-cycle. Performing more relaxations per grid should remove more error, but at a computational cost. The implemented relaxation method is weighted Richardson iteration with a weight, $\omega$, that can be set by the user and defaults to 1. When applied to the transport equation, this is
\begin{equation}
  \phi^{m} = \bigr(\ve{I} + \omega(\ve{TMS} - \ve{I})\bigl)\phi^{m-1} + \omega b^{m-1} \:.
  \label{eq:relax}
\end{equation} 

The MGE preconditioner was designed to take advantage of the energy decomposition used by the MG Krylov method. 
When using multiple energy sets, each set does its own ``mini'' V-cycle with the groups on that set. Each set restricts, prolongs, and relaxes only its own groups and does not need to communicate with other energy sets beyond what is needed for handling upscattering in the solver. This strategy is preferable to doing a full V-cycle across sets, which would require much more data transfer and accounting.
This model is also a communication savings compared to using grids in space or angle. An additional benefit is the simplicity of energy grids. 
Energy is one-dimensional, which allows for simpler coarsening and refinement than spatial or angular grids.
%In the case of an unequal number of groups per set, all sets are forced to have the same grid depth to enforce energy load balancing between sets. Thus, each set restricts to one or two groups, giving approximately $num\_sets$ total groups across all sets at the coarsest level. The number of grids needed is determined by the set with the minimum number of groups, since it will be the first to reach a grid with one group:%. This modifies Equation~\eqref{eq:NumGrids} to be
%\begin{align}
%  num\_g_{min} &= \text{floor}\bigl(\frac{num\_groups}{num\_sets}\bigr) \:, \\
%  num\_grids &= \text{floor}\bigl( \log_{2}(num\_g_{min}) \bigr) + 2 \:.
%  \label{eq:multisetGrids}
%\end{align}

%%---------------------------------------------------------------------------%%
%%---------------------------------------------------------------------------%%
%%---------------------------------------------------------------------------%%
\section{Past Work}
\label{sec:pastwork}
This section summarizes results from using RQI \textit{without} preconditioning \cite{Slaybaugh2012} and the MGE preconditioner used with fixed source problems or PI \cite{Slaybaugh2013}. The purpose of this section is to highlight the capabilities that have been demonstrated, and point out the short-comings that can be overcome by using the MG Krylov solver, RQI eigenvalue solver, and MGE preconditioner together. The goal of using this system of solvers is to improve convergence behavior of the multigroup Krylov solves integrated over eigenvalue solves. The best metric for measuring this is the total number of multigroup Krylov iterations used in a calculation because %it is the most consistent and fair measure. The number of eigenvalue iterations is also compared for the eigenvalue tests--a point of interest rather than a measure of goal attainment. 
%The total number of Krylov iterations is the best proxy for convergence behavior as 
it encompasses the total work done. Timing comparisons should be considered heuristically unless otherwise specified.

\subsection{Unpreconditioned RQI} \label{subsec:rqiresults}
The goal of the RQI studies that have been published was to find if RQI is useful without preconditioning. We solved several problems, where the first set was two small eigenvalue test problems, one with vacuum boundaries and one with reflecting, that had small dominance ratios. We found that RQI got the correct answer and converged in fewer iterations than PI.

However, an intermediately-sized problem did not work so well: the eigenvector did not converge after the first iteration. The value of $k$ oscillated between 0.3966 and 0.3967 (the correct value was 0.4) until the calculation was manually terminated. In the 2-D and 3-D C5G7 MOX Benchmark problems (\cite{OECD-NEA2003}, \cite{OECD-NEA2005}), which are more realistic calculations, RQI did not converge the eigenvector nor find an eigenvalue close to the correct one. 
These cases of RQI's non-convergence are caused by the non-convergence of the inner linear solve. Given a large enough subspace and/or enough iterations (which is not feasible in practice because of memory limitations), the linear solver would always converge and so would RQI. 

The more challenging problems showed that, as expected, the Krylov solver often cannot converge the eigenvector with the ill-conditioned systems created by RQI in a reasonable amount of time. Work by Hamilton \cite{Hamilton2010} also demonstrated the need for preconditioning in solving shifted transport problems.%When the Krylov iterations do not converge, the flux estimate is not good. Without an adequate approximation to the eigenvector, the RQ is no longer a valid approximation to the eigenvalue, and therefore the eigenvalue problem does not converge. 
The small problems, however, showed that RQI can require fewer Krylov iterations than PI if the multigroup iterations are converged. Thus, if the MG Krylov solver is preconditioned so that the eigenvector converges, RQI may be able to find the correct eigenvalue more efficiently than PI for cases of interest. 

This leads to the question: what preconditioner should we choose? Iterative methods reduce oscillatory but not smooth error modes effectively and the smooth error can prevent iterative methods from converging. This behavior is characterized by rapid error reduction in the first several iterations followed by very little error reduction. Such a trend was observed in the tests where RQI failed. Multigrid methods selectively remove smoother error components and are therefore ideal for accelerating this type of problem. 

\subsection{MGE Preconditioner}\label{subsec:mgeresults}
Much of the previously-published work for the MGE preconditioner focused on choosing all of the options that control the preconditioner: Richardson iteration weight, number of V-cycles per preconditioner application, number of relaxations per level, quadrature in the preconditioner, effects of energy sets, and depth of the V-cycle (Hamilton's work \cite{Hamilton2010} also discusses multigrid-cycle parameter selection). The syntax used throughout this work will be that $w\#$ is the weight, $r\#$ is the number of relaxations per level, and $v\#$ is the number of V-cycles, e.g.\ $w1r1v1$ is one relaxation per level, one V-cycle, and a weight of 1.0. Using more preconditioning means using larger values of $w$ and/or $r$ and/or $v$.

Some initial tests provided a basis for choosing each of these. %The results showed that increasing $r$ and/or $v$ decreased Krylov iteration count, as expected. They also indicated that using a small amount of weight can be beneficial, but a large amount is not. 
As a result, the default parameters are $w1r2v2$. Tests also showed that using a reduced quadrature set inside the preconditioner is very valuable. %For example, using $S_2$ inside MGE for an iron-graphite test solved with $S_8$ reduced the solve time by 73\% compared to using $S_8$ in the preconditioner. 
Another important area of investigation demonstrated that the preconditioner scaled very well with multiple energy sets. This was largely because increasing sets reduces the cost of the preconditioner since each set uses a shallower V-cycle and therefore does less work. 
%The problems solved scaled about 40\% more efficiently with MGE preconditioning than without it. The main reason for the preconditioner’s good scaling is that as the number of sets increases, each application of the preconditioner becomes less costly since the V-cycle becomes shallower and therefore each application of the preconditioner performs fewer total relaxations. These results prompted an investigation of controlling V-cycle depth explicitly. 
The results of several tests additionally demonstrated that, in general, it is better to use only a few grids (shallower V-cycle). %The optimal number of grids will be problem dependent, but a default grid depth of two was recommended.

All of these tests inform the best way to use the MGE preconditioner, but do not determine whether and when it is useful. To begin investigating this question, two eigenvalue problems were solved with preconditioned and unpreconditioned PI. The first problem was the 3-D C5G7 MOX Benchmark problem. In this case the use of MGE significantly reduced Krylov iteration count, but increased the overall runtime. A full PWR problem (described in subsection \ref{subsec:PWR}) exhibited similar behavior: fewer iterations in more time. 

These results suggest that the MGE preconditioner was a failed experiment after all. However, the two eigenvalue problems solved were only solved with PI and the mathematical properties of the MGE preconditioner suggest it would benefit the RQI solver. Thus, the work published so far has not settled the question of whether MGE is a useful preconditioner for at least some problems. 

%%---------------------------------------------------------------------------%%
%%---------------------------------------------------------------------------%%
%%---------------------------------------------------------------------------%%
\section{Results} \label{sec:results}
The collection of observations in the past work section led to the questions (1) Will preconditioning with MGE facilitate the use of RQI? and (2) Will the combination of RQI, MGE, and the block Krylov solver be advantageous for at least some problems? This section is designed to answer these two questions. In this Section, reducing the Krylov iteration count and reducing total compute time are measures of success. 
Unless otherwise noted, all test problems used a step characteristic spatial solver, level-symmetric angular quadrature, and the grid depth was determined using the default approach.
The Krylov solver was GMRES($m$), which is set to limit the number of multigroup iterations to 1,000 if the problem does not converge earlier. The convergence tolerances are noted for each problem. The tolerance for the multigroup solve is the convergence tolerance used by GMRES in Trilinos. The eigenvalue tolerance is used by PI or RQI to determine if the eigenvalue has converged. %In Denovo, PI also checks the L2-norm and the infinity-norm of the difference in the fission source between iterations. The default L2-norm tolerance is 1.0 and infinity-norm tolerance is 0.01.

%%---------------------------------------------------------------------------%%
\subsection{2-D C5G7} \label{subsec:2dc5g7}
%A variety of small problems were solved to verify code functionality, ensure that total Krylov count is reduced by the addition of MGE, verify MGE parameter choice when used in combination with RQI, and confirm that the preconditioner should \emph{not} use the shifted version of the operator.
%
%The first test problem set was that initially solved with unpreconditioned RQI mentioned in Section~\ref{subsec:rqiresults}, which included a vacuum and a reflecting boundary case. The addition of MGE to the vacuum test always resulted in the correct solution in fewer Krylov iterations than the unpreconditioned case. The number of RQ iterations was not affected by the preconditioning. Adding MGE in the reflecting case reduced both total Krylov iteration count and eigenvalue iteration count, except for cases in which $r1v1$ was used with a weight of 1.2 or higher. Larger $r/v$ combinations did not have difficulty with a weight of 1.2. These results confirm that the solvers can work together, that the correct solution is obtained as long as the problem converges, that Krylov iteration count is reduced when RQI is used with MGE, and the parameter guidance previously established is still valid. 
%
Using guidance from the past work, MGE was applied to the 2-D C5G7
benchmark using both PI and RQI. The goals were to see if
preconditioned RQI could converge the eigenvector (flux) and solve for $k$, and to
investigate the effect of preconditioning with both RQI and PI.
%, and to see whether the lessons learned about preconditioning parameters still hold in a real problem. 
The calculation used 16 cores on the small Orthanc cluster at Oak Ridge: 4 $x$-blocks, 4 $y$-blocks, 1 $z$-block and 1 energy set. The total and upscattering tolerances were 1 $\times$ 10$^{-3}$, with a $k$ tolerance of 1 $\times$ 10$^{-5}$. %An optimized version of Denovo was used. 

The first study used the MGE preconditioner with PI. The results are shown in Table~\ref{table:2-D c5g7}, where ``Krylov'' is the total number of Krylov iterations and ``PI'' is the total number of eigenvalue iterations.
All calculated $k$s were within the uncertainty of the benchmark and so are not reported.
This study shows the preconditioner is very effective at reducing the number of Krylov iterations used by PI, but PI is much faster without preconditioning. The unpreconditioned case, corresponding to a weight of 0, took about twice the MG Krylov iterations as the $w\#r1v1$ cases. Adding the preconditioner dramatically reduced the number of Krylov iterations but more than doubled the runtime. With $r1v1$, increasing the weight from 1 to 1.4 decreased the number of Krylov iterations and the time to solution. The time and iteration count both increased with a weight of 1.5. When the weight was increased beyond 1.5 none of the multigroup iterations converged so they are not reported here. The two calculations with a higher level of preconditioning converged in the fewest Krylov iterations, with the $w1.4r2v2$ taking the smallest amount of time and $w1r3v3$ taking the second largest.
\begin{table}[h!]
  \centering
  \caption{2-D C5G7 benchmark: preconditioner weight variation with PI}
  \begin{tabular}{c c c c c c}
    \toprule
    Weight & Relaxations & V-cycles & Krylov & PI & Time (s) \\
    \midrule
    0.0 & 0 & 0 & 3,129 & 32 & 8.54 $\times$ 10$^{3}$ \\
    1.0 & 1 & 1 & 1,649 & 31 & 2.12 $\times$ 10$^{4}$ \\
    1.1 & 1 & 1 & 1,591 & 31 & 2.06 $\times$ 10$^{4}$ \\
    1.2 & 1 & 1 & 1,546 & 31 & 1.98 $\times$ 10$^{4}$ \\
    1.3 & 1 & 1 & 1,492 & 31 & 1.92 $\times$ 10$^{4}$ \\
    1.4 & 1 & 1 & 1,458 & 31 & 1.91 $\times$ 10$^{4}$ \\
    1.5 & 1 & 1 & 1,771 & 31 & 2.29 $\times$ 10$^{4}$ \\
    1.4 & 2 & 2 & 438   & 31 & 1.77 $\times$ 10$^{4}$ \\
    1.0 & 3 & 3 & 253   & 31 & 2.28 $\times$ 10$^{4}$ \\
    \bottomrule 
  \end{tabular} 
  \label{table:2-D c5g7}
\end{table}
%\begin{figure}[!ht]
%    \begin{center}
%      \includegraphics [width=0.6\textwidth, height=0.6\textheight] {2Dc5g7PI}
%   \end{center}
%   \caption{2-D C5G7 Benchmark, Preconditioner Weight Variation with Power Iteration}
%   \label{fig:2-Dc5g7PI}
%\end{figure}

The results from the preconditioned RQI study are in Table~\ref{table:2-D c5g7 rqi}. In all cases, except the unpreconditioned one, $k$ was within the uncertainty of the benchmark value. The ``$<$ 1,000?'' column indicates whether or not the multigroup iterations converged during the RQI process, %If the value is ``no,'' the eigenvector only converged during the first iteration. 
where a number indicates the last eigenvalue iteration for which the Krylov method took less than 1,000 iterations. % and all subsequent iterations required the full 1,000. A ``yes'' means all of the Krylov iterations converged. 
The relative time is the ratio of the case of interest to the unpreconditioned PI time of 8.54 $\times$ 10$^{3}$ seconds.
\begin{table}[h!]
  \centering
  \caption{2-D C5G7 benchmark: convergence study with RQI}
  \begin{tabular}{c  c  c  l  c  c  c }
    \toprule
    Weight & Relaxations & V-cycles & Krylov & RQI & $<$ 1,000? & Rel.\ Time\\ %$^{\dag}$\\
    \midrule
    0   & 0 & 0 & 119,006 & 120$^{*}$ & no  & 10.98 \\%$9.38 \times 10^{4}$ \\
    1   & 1 & 1 & 16,007  & 17        & no  & 23.65 \\ %$2.02 \times 10^{5}$ \\
    1.2 & 1 & 1 & 40,008  & 41$^{*}$  & no  & 13.00 \\ %$2.06 \times 10^{5}$ \\
    1   & 3 & 1 & n/a     & n/a$^{*}$ & 7   & n/a \\
    1   & 2 & 2 & 11,158  & 19        & alternated & 46.72 \\ %$3.99 \times 10^{5}$ \\
    1   & 3 & 2 & 3,320   & 19        & 14  & 19.23 \\ % $1.64 \times 10^{5}$ \\
    %\hline
    1   & 3 & 3 & 299     & 19        & yes & 3.01 \\ %$2.57 \times 10^{4}$ \\
    1.1 & 3 & 3 & 281     & 19        & yes & 2.80 \\ %$2.40 \times 10^{4}$ \\
    1.3 & 3 & 3 & 254     & 19        & yes & 2.57 \\ %$2.19 \times 10^{4}$ \\
    1.5 & 3 & 3 & n/a     & n/a$^{*}$ & no  & n/a \\
    \bottomrule 
  \end{tabular} \\
  %$^{\dag}$compared to unpreconditioned PI, $8.54 \times 10^{3}$ seconds\\
  $^{*}$terminated manually
  \label{table:2-D c5g7 rqi}
\end{table}

These results show a few important things. Most significantly, with enough preconditioning the multigroup iterations within RQI can be converged and the correct eigenpair can be found. This was true even when the eigenvector did not converge inside each eigen iteration (though these cases were significantly more time consuming). Further, as the preconditioning increased the eigenvector came closer to converging for all iterations. For the first three $w\#r3v3$ cases, all of the Krylov iterations converged. This test case was the first to demonstrate that the preconditioner can get RQI to converge when it did not converge without preconditioning. When the Krylov iterations did converge, increasing the weight decreased iteration count and wall time for small weights. Finally, too much weight prevented the calculation from converging.

A big question is whether RQI is faster than PI with preconditioning, but only the $w1r3v3$ calculation overlapped between RQI and PI. PI took fewer Krylov iterations, 253 compared to 299, and less time, 2.28 $\times$ 10$^{4}$ compared to 2.57 $\times$ 10$^{4}$ seconds. From the standpoint of comparing eigenvalue solution methods, it is worth noting that RQI required 19 eigenvalue iterations while PI required 31. Thus, when given eigenvectors that have been converged to the same tolerance, RQI needed fewer eigenvalue iterations than PI. For this test preconditioned RQI did not perform as well as preconditioned PI, though the times and iteration counts were close to one another. 

\subsection{3-D C5G7} \label{subsec:3dc5g7}
Next, the preconditioner using both PI and RQI was applied to the 3-D C5G7 benchmark. %with an optimized version of Denovo. 
The goals of this study were essentially the same as the 2-D study, except that this problem is larger and represents a more realistic problem. The medium-sized OIC cluster at Oak Ridge was used, and each problem was given 720 cores with 40 $x$-blocks, 18 $y$-blocks, and 5 $z$-blocks. The total and upscattering tolerances were 1 $\times$ 10$^{-4}$, with a $k$ tolerance of 1 $\times$ 10$^{-5}$ unless otherwise indicated. The wall time limit was 12 hours. %Note that on the OIC machine if a calculation exceeds wall time there is no way to get any of the results from the scratch space. 

The preconditioned PI results were reported in \cite{Slaybaugh2013}, so we only add the RQI results shown in Table~\ref{table:3-D c5g7 rqi} here. The relative time is compared to unpreconditioned PI, 4.46 $\times$ 10$^{3}$ seconds. Cases with no or a small amount of preconditioning ($w1r1v1$, $w1.5r1v1$, $w1.2r2v1$) are not reported since none converged within the walltime limit. This indicates a small amount of preconditioning was insufficient. However, with a lot of preconditioning ($w1r4v4$, $w1.5r5v5$) too much time was spent in the preconditioner and the problem did not finish in time either. Convergence was obtained in several cases by lowering the tolerances.
\begin{table}[h!]
  \centering
  \caption{3-D C5G7 benchmark: preconditioner parameter scoping with RQI}
  \begin{tabular}{ c  c  c  c  l  c }
    \toprule
    Weight & Relaxations & V-cycles & Krylov & RQI & Rel.\ Time\\%$^{\dag}$ \\[0.5ex]
    \midrule
    %0   & 0 & 0 & n/a & n/a      & exceeded wall time \\
    %1   & 1 & 1 & n/a & n/a      & exceeded wall time \\
    %1.5 & 1 & 1 & n/a & n/a      & exceeded wall time \\
    %1.2 & 2 & 1 & n/a & n/a      & exceeded wall time \\
    1.3 & 2 & 2 & 302 & 19       & 5.20 \\ %$2.32 \times 10^{4}$ \\
    1   & 3 & 3 & 103 & 9$^{*}$  & 6.67 \\ %$3.02 \times 10^{4}$ \\
    1   & 3 & 3 & 164 & 15$^{+}$ & 7.59 \\ %$3.38 \times 10^{4}$ \\
    1.5 & 3 & 3 & 187 & 19       & 7.26 \\ %$3.24 \times 10^{4}$ \\
    1   & 4 & 4 & n/a & n/a      & exceeded wall time \\
    1   & 4 & 4 & 74  & 9$^{*}$  & 5.13 \\ %$2.29 \times 10^{4}$ \\
    1.5 & 5 & 5 & n/a & n/a      & exceeded wall time \\
    \bottomrule 
  \end{tabular}\\
  %$^{\dag}$compared to unpreconditioned PI, 4.46 $\times$ 10$^{3}$ seconds\\
  $^{*}$tol and upscatter tol = 1 $\times$ 10$^{-5}$, $k$ tol = 1 $\times$ 10$^{-3}$\\
  $^{+}$tol and upscatter tol = 1 $\times$ 10$^{-4}$, $k$ tol = 5 $\times$ 10$^{-5}$
  \label{table:3-D c5g7 rqi}
\end{table}  

With an intermediate amount of preconditioning, RQI converged and performed better than the analogous PI cases. There are three cases where both problems finished and the same tolerances were used: $w1.3r2v2$, $w1r3v3$, $w1.5r3v3$. These results are shown together in Table~\ref{table:PI RQI}. This table displays time instead of relative time to facilitate comparison between two cases rather than across all cases. In all three pairs, the RQI calculations took less time and fewer eigenvalue iterations. %In the second two, they also took fewer Krylov iterations. %RQI even finished in time to get results from the $w1r4v4$ calculation when PI did not. 
\begin{table}[h!]
  \centering
  \caption{3-D C5G7 benchmark: RQI and PI comparison}
  \begin{tabular}{ c  c  c  c  c  c  c }
    \toprule
    Sovler & Weight & Relaxations & V-cycles & Krylov & Eigenvalue & Time (s) \\[0.5ex]
    \midrule
    RQI & 1.3 & 2 & 2 & 302    & 19       & 2.32 $\times$ 10$^{4}$ \\
    PI  & 1.3 & 2 & 2 & 288    & 32       & 2.84 $\times$ 10$^{4}$ \\
    %\hline
    RQI & 1   & 3 & 3 & 103    & 9$^{*}$  & 3.02 $\times$ 10$^{4}$ \\
    PI  & 1   & 3 & 3 & 126    & 14$^{*}$ & 4.04 $\times$ 10$^{4}$ \\
    %\hline
    RQI & 1.5 & 3 & 3 & 187    & 19       & 3.24 $\times$ 10$^{4}$ \\
    PI  & 1.5 & 3 & 3 & 192    & 32       & 3.73 $\times$ 10$^{4}$ \\
    \bottomrule 
  \end{tabular}\\
  $^{*}$tol and upscatter tol = 1 $\times$ 10$^{-5}$, $k$ tol = 1 $\times$ 10$^{-3}$
  \label{table:PI RQI}
\end{table}  

The 3-D benchmark problem shows that for at least some problems, preconditioned RQI converges more quickly in all senses than preconditioned PI if a sufficient amount of preconditioning is used to get RQI to converge. It is pertinent that this is true in the most interesting problem shown so far. However, unpreconditioned PI is still superior in terms of timing.%It seems, however, that RQI can only be useful if it is preconditioned enough to get the eigenvector to converge. 
%** Could do more of these with better w,r,v choices to have more overlap between RQI and POI; use reduced angle set; not much grid depth to reduce, but could try anyway **

\subsection{PWR 900} \label{subsec:PWR}
Finally, the preconditioner using both PI and RQI was applied to a very challenging problem--a detailed PWR. This problem had 44-groups; 578 $\times$ 578 $\times$ 700 mesh elements (233,858,800 cells) broken over 112 $\times$ 112 $\times$ 10 partitions (12,544 blocks); used a $P_0$ scattering expansion; used an $S_{12}$ angular quadrature, with $S_2$ in the preconditioner; and was restricted to a V-cycle depth of 2. The total number of unknowns was thus 1.73 trillion. Based on PWR calculations done previously by Evans and Davidson \cite{Evans2010}, $k$ is approximately 1.27. The convergence tolerance was 1 $\times$ 10$^{-3}$ and the $k$ tolerance = 1 $\times$ 10$^{-3}$. This was solved on Titan. 

The results using 11 energy sets are given in Table~\ref{tab:PWR all} and show that RQI was much faster and required far fewer Krylov and eigenvalue iterations than PI for this problem. In fact, The preconditioned PI problems did not finish before the wall time limit was reached (RQI was only run with preconditioning since other tests indicated it would not converge within a reasonable time without it). This is the first case where preconditioned RQI was better on all counts than PI with or without preconditioning. RQI with MGE was more than 10 times faster than PI. %This test shows that RQI can be the better eigenvalue solver for at least some problems. It is promising that the calculation for which RQI is decisively faster than PI is the one for which this work was designed. 
\begin{table}[h!]
  \centering
  \caption{PWR-900: PI and RQI with and without preconditioning, 11 energy sets}
    \begin{tabular}{ c  c  c  c  c  c  c }
      \toprule
      Method & Precond & N Eigen & N Krylov & $k$ & Time (m) \\
      \midrule
      %RQI & none   & n/a & n/a   & n/a                                & n/a$^*$ \\
      %RQI & w1r1v1 & 3   & 45    & 1.252 $\pm$ 1.42 $\times$ 10$^{-2}$ & 120$^*$ \\
      RQI & w1r2v2 & 5   & 70    & 1.268 $\pm$ 1.24 $\times$ 10$^{-3}$ & 54.8 \\
      RQI & w1r3v3 & 6   & 76    & 1.269 $\pm$ 1.12 $\times$ 10$^{-3}$ & 330.4$^{\dag}$ \\
      PI  & none   & 149 & 5,602 & 1.276 $\pm$ 1.85 $\times$ 10$^{-3}$ & 612.2 \\
      %PI  & w1r1v1 & info &      &                     & \\
      PI  & w1r2v2 & 86  & 946   & 1.275 $\pm$ 1.43 $\times$ 10$^{-3}$ & 720$^*$ \\
      PI  & w1r3v3 & 11  & 111   & 1.270 $\pm$ 5.09 $\times$ 10$^{-2}$ & 480$^{*,\dag}$ \\
      %\hline
      %PI  & none   & 24  & 901  & 1.272 $\pm$ 1.01e-4 & 92.1$^+$ \\
      %PI  & w1r2v2 & 28  & 312  & 1.273 $\pm$ 8.24e-5 & 250.9$^+$ \\
      \bottomrule
    \end{tabular}\\
    $^{*}$did not converge within walltime limit (8 or 12 hours)\\
    $^{\dag}$used $S_{12}$ in MGE preconditioner; full V-cycle depth;
    tolerance = 1 $\times$ 10$^{-4}$, $k$ tolerance = 1 $\times$ 10$^{-3}$; 
    102 $x$-blocks, 100 $y$-blocks, and 10 $z$-blocks\\
  \label{tab:PWR all}
\end{table}

To investigate how the full system performs for a real problem, a strong scaling study using RQI with MGE using $w1r2v2$ and 1, 4, 11, and 22 sets was done. 
The results are given in Table~\ref{tab:PWR rqi strong scaling} where $\text{t}_{\text{perfect}}$ = ($\text{1 set solve time}$ / $\text{\# energy sets}$) and efficiency = ($\text{t}_{\text{perfect}}$ / $\text{t}_{\text{actual}}$). 
A strong scaling study with MGE has been published before, but in that case the V-cycle depth was not fixed.
This meant that increasing energy sets decreased V-cycle depth such that the preconditioner did less work with more sets. 
In this study the amount of work done by the preconditioner does not vary with energy sets.
\begin{table}[h!]
  \centering
  \caption{PWR-900: RQI strong scaling with $w1r2v2$ preconditioning}
  \begin{tabular}{ c  c  c  c  c  c  c }
     \toprule
      Sets & Domains & Time (m) & $\text{t}_{\text{perfect}}$ & Efficiency \\
      \midrule
      1   & 12,544 & 407.8 & 407.8 & 1.000 \\
      4   & 50,176 & 123.4 & 102.0 & 0.826\\
      11  & 137,984 & 54.8 & 37.1  & 0.676\\
      22  & 275,968 & 39.6 & 18.5  & 0.468\\
      \bottomrule
  \end{tabular}
  \label{tab:PWR rqi strong scaling}
\end{table}

The scaling compares quite well to previous scaling studies for Denovo. A fixed source (i.e.\ MG Krylov only) problem with a similar mesh and 44 groups scaled from 4,320 domains to 190,080 domains with an efficiency of 0.64 \cite{Slaybaugh2011}. That this problem performed similarly shows that adding RQI and the MGE preconditioner as solvers does not degrade the strong scaling achieved using the MG Krylov solver only. It is promising that the new solver system does not degrade scaling and that the calculation for which RQI is decisively faster than PI is the one for which this work was designed. 

%%---------------------------------------------------------------------------%%

\section{Conclusions}
\label{sec:conclusions}
The goal of this research was to accelerate transport calculations with methods that can take full advantage of modern leadership-class computers, facilitating the design of better nuclear systems. Three complimentary methods were used together to demonstrate improvement over traditional methods: the MG Krylov solver, RQI, and the MGE preconditioner. 
The multigroup Krylov solver converges more quickly than GS and enables energy decomposition such that Denovo can scale to hundreds of thousands of cores. The new multigrid in energy preconditioner reduces iteration count for many problem types and takes advantage of the new energy decomposition such that it can scale very efficiently. These two tools are useful on their own, but together they allow the Rayleigh quotient iteration eigenvalue solver to work. These ideas have never before been used together in this way.  

The real motivation of this work was to add RQI, which should converge in fewer iterations than PI for large and challenging problems. RQI creates shifted systems that would not be tractable without the MG Krylov solver. It also creates ill-conditioned matrices that cannot converge without the multigrid in energy preconditioner. Using these methods, RQI converged in fewer iterations and in less time than all PI calculations for a full PWR core. It also scales reasonably well out to 275,968 cores. 

The methods added in this research accelerated Denovo in multiple ways. This acceleration helps enable the solution of today's ``grand challenge" problems. It is hoped that improved methods will lead to improved reactor designs and systems, and that the frontier of computational challenges will be moved forward.

%%---------------------------------------------------------------------------%%

\section{Acknowledgements}

This research used resources of the Oak Ridge Leadership Computing Facility at the Oak Ridge National Laboratory, which is supported by the Office of Science of the U.S. Department of Energy under Contract No. DE-AC05-00OR22725. Additional thanks to the Rickover Fellowship Program in Nuclear Engineering sponsored by Naval Reactors Division of the U.S. Department of Energy. This fellowship sponsored the work from which this work is derived. 

%%%%%%%%%%%%%%%%%%%%%%%%%%%%%%%%%%%%%%%%%%%%%%%%%%%%%%%%%%%%%%%%%%%%%
\setlength{\baselineskip}{12pt}

\bibliographystyle{mc2015}
\bibliography{RQI_MGE}

\end{document}